# A GPU implementation of the Simulated Annealing Heuristic for the Quadratic Assignment Problem

Gerald Paul

*Center for Polymer Studies and Dept. of Physics, Boston University*

*Boston, Massachusetts 02215*

**Abstract**

The quadratic assignment problem (QAP) is one of the most difficult combinatorial optimization problems. An effective heuristic for obtaining approximate solutions to the QAP is simulated annealing (SA). Here we describe an SA implementation for the QAP which runs on a graphics processing unit (GPU). GPUs are composed of low cost commodity graphics chips which in combination provide a powerful platform for general purpose parallel computing. For SA runs with large numbers of iterations, we find performance $\approx 50 - 100$ times better than that of a recent non-parallel but very efficient implementation of SA for the QAP.

## 1. Introduction

Recently, relatively inexpensive commodity graphics chips produced in large volumes to provide basic graphics support on personal computers have been combined into powerful graphics processing units (GPUs) which can be used for general purpose parallel processing [9]. Here we describe the implementation of the simulated annealing heuristic on such a GPU.

Originally formulated by Koopmans and Beckmann [12], the QAP is NP-hard and is considered to be one of the most difficult problems to be solved optimally. It was defined in the following context: A set of $N$ facilities are to be located at $N$ locations. The quantity of materials which flow between facilities $i$ and $j$ is $A_{ij}$ and the distance between locations $i$ and $j$ is $B_{ij}$. The problem is to assign to each location a single facility so as to minimize the cost

$$C = \sum_{i=1}^{N} \sum_{j=1}^{N} A_{i,j} B_{p(i),p(j)}, \qquad (1)$$

where $p(i)$ represents the location to which $i$ is assigned.

There is an extensive literature that addresses the QAP and which is reviewed in Refs. [1, 6, 10, 13, 15, 4]. The QAP is exceedingly hard to solve optimally. With the exception of specially constructed cases, optimal algorithms have solved only relatively small instances with $N \leq 36$. Therefore, various heuristic approaches have been developed and applied to problems typically of size $N \approx 100$ or less. In contrast, a travelling salesman problem consisting of almost 25,000 towns in Sweden has been solved exactly [2].

## 2. Background

### 2.1. Simulated Annealing

The simulated annealing heuristic was first applied to the QAP by Burkhard and Rendl [5] and was refined by Connolly [7]. The heuristic consists of swapping locations of two facilities. Proposed swaps can either be determined randomly or selected according to some sequential enumeration of all possible swaps. For each proposed swap, the change in cost, $\delta$, for the potential swap is calculated. The swap is made if

$$\begin{aligned} \delta &< 0 \quad or \\ e^{-\delta/T} &> r, \end{aligned} \qquad (2)$$

where $T$ is an analog of temperature in physical systems that is slowly decreased according to a specified *cooling schedule* after each iteration and $r$ is a uniformly distributed random variable between 0 and 1. Randomly making moves which increase the cost is done to help escape from local minima.

In traditional implementations of the heuristic, the cost of making a swap is calculated from scratch when the swap is considered in order to determine if the swap should be made.

## 2.2. Recent efficient non-parallel implementation

Recently an implementation of SA for the QAP was developed which improves performance over the traditional implementation by factors of up to 100 for large numbers of iterations on large problem instances [17]. The approach taken is to attempt to reduce the complexity by maintaining a matrix $\Delta$ of costs of each possible swap. This approach is motivated by the work of Taillard [18]), who applied a similar approach in the application of another heuristic, tabu search, to the QAP. The $\Delta$-*matrix approach* is as follows:

(a) Initialize by creating a matrix $\Delta_{i,j}$ containing the cost of swapping $i$ and $j$ for all $i$ and $j$, given a current assignment $p$. Set iteration number to zero.

(b) Increment the iteration number. Retrieve the cost $\Delta_{r,s}$ of the next possible swap $(r, s)$.

(c) If the cost of the possible swap does not meet the criteria in Eq. (2), go to (b).

(d) If the proposed swap meets the swap criteria, update $p$ to reflect the swap, update $\Delta_{i,j}$ with the new swap costs given the new permutation, and go to (b).

(e) End after $I$ iterations.

The number of iterations in which a swap is performed divided by the total number of iterations is known as the *acceptance rate* $a(I)$. The lower the acceptance range, the more efficient the $\Delta$-matrix approach.

In addition to providing higher performance, even in a non-parallel environment, than the traditional SA implementation, the $\Delta$-matrix implementation lends itself to GPU processing because, as explained in Section 3, the costs of all swaps are calculated together at the same time.

## 2.3. Need for parallelization

Ideally if many processors are available to run a heuristic, the most efficient use of these processors is to ran copies of the heuristic independently on each processor. However, as shown in [17], statistically, the larger the number of SA iterations, the higher the quality of the solution found. Thus if high quality solutions are desired, an SA run with a high number of iterations can not be replaced by many independent SA runs each with a smaller number of iterations; a parallel implementation is therefore required to allow timely completion of SA runs with a high number of iterations.

## 2.4. Previous work on parallel heuristics for the QAP

James et al. [11] propose a parallel tabu search algorithm for the QAP and review previously proposed parallel heuristics for the QAP most of which involve the tabu search heuristic. Recently it was shown that if high quality solutions are desired, simulated annealing performs better than tabu search [16]. We are not aware of any implementations of the simulated annealing heuristic for the QAP on a GPU. Simulated annealing presents a complexity not present in tabu search. Instead of calculating all possible swaps and then choosing one, a swap is made when the first swap meeting the requirement of Eq. (2) is found.

Crainic and Toulous [8] classify parallel heuristic methods into three categories. Our implementation falls into their Type 1 category — the parallelization of computationally expensive low level operations.

## 2.5. Characteristics of graphics processing unit

We have implemented the SA heuristics on the Nvidia C2070 GPU. The important features of the NVIDIA GPU architecture are [14]:

- A graphics processing unit contains one or more symmetric multiprocessors(SMPs). Each SMP contains multiple processing elements each of which executes a sequential *thread*. Threads are organized into *warps* of 32 threads each. One or more warps are further grouped into a *block*.

- Threads within a block can be synchronized with a barrier synchronization GPU primitive.

- Code which runs on the GPU is contained in a *kernel*. There is a relatively large overhead, on the order of microseconds, associated with launching a kernel.

- Each SMP contains a small amount of *shared memory* which all of the processing elements can access with relatively low latency. For the GPU on which we implemented our program shared memory is 48 KB.

- All SMPs can access a large *device memory* but with high latency on the order of 100 times longer than access to shared memory; for the GPU on which we implemented our program device memory is 1.3 TB. Data must be transferred from the host CPU memory to device memory before it can be processed by an SMP.

- When a warp executes an instruction that accesses device memory, it *coalesces* the memory accesses within the warp into one or more memory transactions for fixed length aligned segments(128 bytes for the C2070) in device memory. The greater the coalescence (the fewer the memory transactions) the better the performance.

## 3. Approach

In the $\Delta$-matrix approach, on which we base our GPU implementation, the bulk of the processing time is spent (i) updating the matrix $\Delta$ and (ii) passing through the matrix elements to identify a swap which meets the criteria of Eq. (2). On the GPU we implement these steps in the algorithm as follows:

(i) We update $\Delta$ using multiple threads in multiple blocks. Each thread calculates elements $\Delta_{i,j}$ with a given $i$ and a range of $j$. If the range of $j$ is too small, the setup overhead to process these elements will be relatively large. For this reason, we limit the number of threads that are used to evaluate $\Delta$ with the constraint that each thread processes at least 16 matrix elements.

(ii) The search for the next potential swap which satisfies Eq. (2) is also performed with multiple threads. Starting at the next possible swap after the last swap performed, multiple threads evaluate a subset of all elements of the $\Delta$-matrix. The size of the subset is an adjustable parameter. If one or more valid swaps are found in the subset, the swap which would have been found first if the swaps were evaluated sequentially is chosen. If no valid swap is found, the next subset is treated. The process is repeated until a valid move is found. If the subset size is too small, before a valid swap is found many subsets must be processed with some attendant overhead. If the subset is too large, more swaps than need be are evaluated since many valid swaps will be found. We found good results when the subset size is determined by the requirement that each thread processes 16 matrix elements.

In order to implement the algorithm efficiently on the Nvidia GPU the following must be taken into account:

(i) With the exception of small problem instances ($N \lesssim 100$. The matrices $A, B$ and $\Delta$ are too large to be stored in shared memory and must be stored in device memory. As noted above in Sec. 2.5, efficient access to device memory is achieved when data in contiguous blocks of 128 bytes is accessed simultaneously by all threads in a warp. In updating the $\Delta$-matrix, terms of the form $A_{i,j}B_{p(i),p(j)}$ must be evaluated. Because it is not possible to control the permutation $p$, it is not possible to meet the requirement for efficient access when updating the $\Delta$-matrix without a modification in the approach. For this reason, we maintain a matrix $B'$, the elements of which reflect swaps which have been performed. In conjunction with a swap of facilities $r$ and $s$, after updating $p$ we update the matrix $B'$

$$B'_{i,j} \leftarrow B'_{p(i),p(j)} \qquad (3)$$

for columns and rows of $B'$ with index $r$ or $s$. This is equivalent to swapping rows $r$ and $s$ and swapping columns $r$ and $s$. The operation is of complexity $O(N)$ and is performed efficiently using $N$ threads making an insignificant contribution to the processing time. Using $B'$ instead of $B$ in the updating of $\Delta$, the terms to be evaluated are of the form $A_{i,j}B'_{i,j}$ and the requirement for efficient data access can be met. The insight that the calculation can be transformed in this way is one of the keys to good performance.

(ii) Because all elements of $\Delta$ are updated at the same time, we are able to use the limited amount of shared memory accessible by the SMPs to improve performance of updating $\Delta$ as follows. After a swap of facilities

(iii) $r$ and $s$, but before updating $\Delta$, we load $A_{r,k}, A_{s,k}, B'_{r,k}$, and $B'_{s,k}$ for all $k$ into shared memory requiring only $O(N)$ bytes of shared memory. No other elements of $A$ and $B'$ are are required for the update of elements $\Delta_{i,j}$ for $i$ and $j$ not equal to $r$ or $s$; other elements of $\Delta$ do require some $A$ and $B'$ device memory access.

(iii) There are points in the processing (e.g. after the identification of the next swap, after the updating of $B'$, and after the updating of $\Delta$) at which a given thread must access a data element updated by another thread (possibly in another block). In order to ensure that the data updating is complete before the data is used, a synchronization mechanism is needed. Threads within the same block can be synchronized with the hardware provided barrier synchronization mechanism. No such mechanism is provided for synchronization between threads in different blocks. For this reason, we have implemented the synchronization mechanism for threads in all blocks proposed in [19]. When synchronizing among blocks, there is a potential for deadlock situations. Deadlock can occur when a block is queued for processing on an SMP while another block is processing on that SMP. For this reason, we limit the number of blocks in a kernel to be no greater than the number of SMPs. The alternative of completing all GPU processing when synchronization is required and then re-launching the GPU kernel is not feasible because of the associated kernel launch overhead.

(iv) The highest performance is achieved when as many threads as possible are executing simultaneously. Using the $\Delta$-matrix approach all the swaps costs are updated at the same place in the implementation, providing a straightforward opportunity to use many threads.

## 4. Numerical Results

We perform numerical experiments on a family of random QAP instances created in the same manner as the Taixxa instances [18] in QAPLIB [3]: the $A$ and $B$ matrices are symmetric with zero diagonal; the matrix elements are chosen from independent uniform distributions. These same instances were analyzed in Ref. [17]. Because the GPU version implements the identical heuristic as the non-parallel implementation, the performance improvement is not dependent on the nature of the instances used to measure performance and it is sufficient to compare performance on one family of instances.

The non-parallel version, written in C++ and developed in Ref. [16], was run on an Intel Xeon 2.4 GHz processor. The GPU version was run on a Nvidia Tesla C2070 GPU with processing elements with clock speed of 1.15 GHz. The GPU contains 14 SMPs each of which contains 32 processing elements. The GPU version is written in C++ with Nvidia CUDA language extensions [14].

In Fig. 1, for various problem sizes, we plot run time versus iterations per run, $I$, for the efficient non-parallel implementation and our parallel GPU based implementation. We note that the GPU based implementation becomes more efficient than the non-parallel implementation for $I \approx 10^4$. Until the iteration number becomes $\approx 10^6$, the non-parallel method does not use the $\Delta$-matrix approach because the acceptance rate of swaps is high (see [17] where the role of acceptance rate is discussed in detail). Our parallel implementation uses the delta matrix for all values of $I$ and therefore suffers when the acceptance rate is high as is the case for low values of $I$. This effect is also seen in Fig. 2 in which we plot the performance improvement

$$P = \frac{t_{non-parallel}}{t_{parallel}} \qquad (4)$$

versus $I$ where $t_{non-parallel}$ and $t_{parallel}$ are the run times for the non-parallel and parallel implementations respectively. For all problem sizes studied, $P$ increases with increasing number of iterations until about $I \approx 10^7$ at which point $P$ reaches a plateau; the additional time contribution from the initial $10^6$ iterations in the parallel implementation then becomes negligible.

## 5. Discussion and Summary

Because

- our parallel implementation provides performance improvements of up to a factor of 100 versus the non-parallel implementation of the $\Delta$-matrix approach, and

- the non-parallel $\Delta$-matrix approach provides a performance improvement of up a factor of 100 versus the traditional implementation [17], the current parallel implementation provides a remarkable performance improvement of up to 10,000 versus the traditional implementation which has been in use essentially unchanged for over two decades.

We emphasize that we make no change to the original SA heuristic to accommodate the GPU environment — only to its implementation. Thus our GPU implementation of the heuristic provides exactly the same results, statistically, as does the traditionally implemented version.

For more details on the GPU implementation we refer the reader to C++/CUDA code provided as supplementary material.

## Acknowledgments

We thank Dan Kamalic for technical assistance in using Boston University's eng-grid GPU enabled computer systems and the Defense Threat Reduction Agency (DTRA) for support.

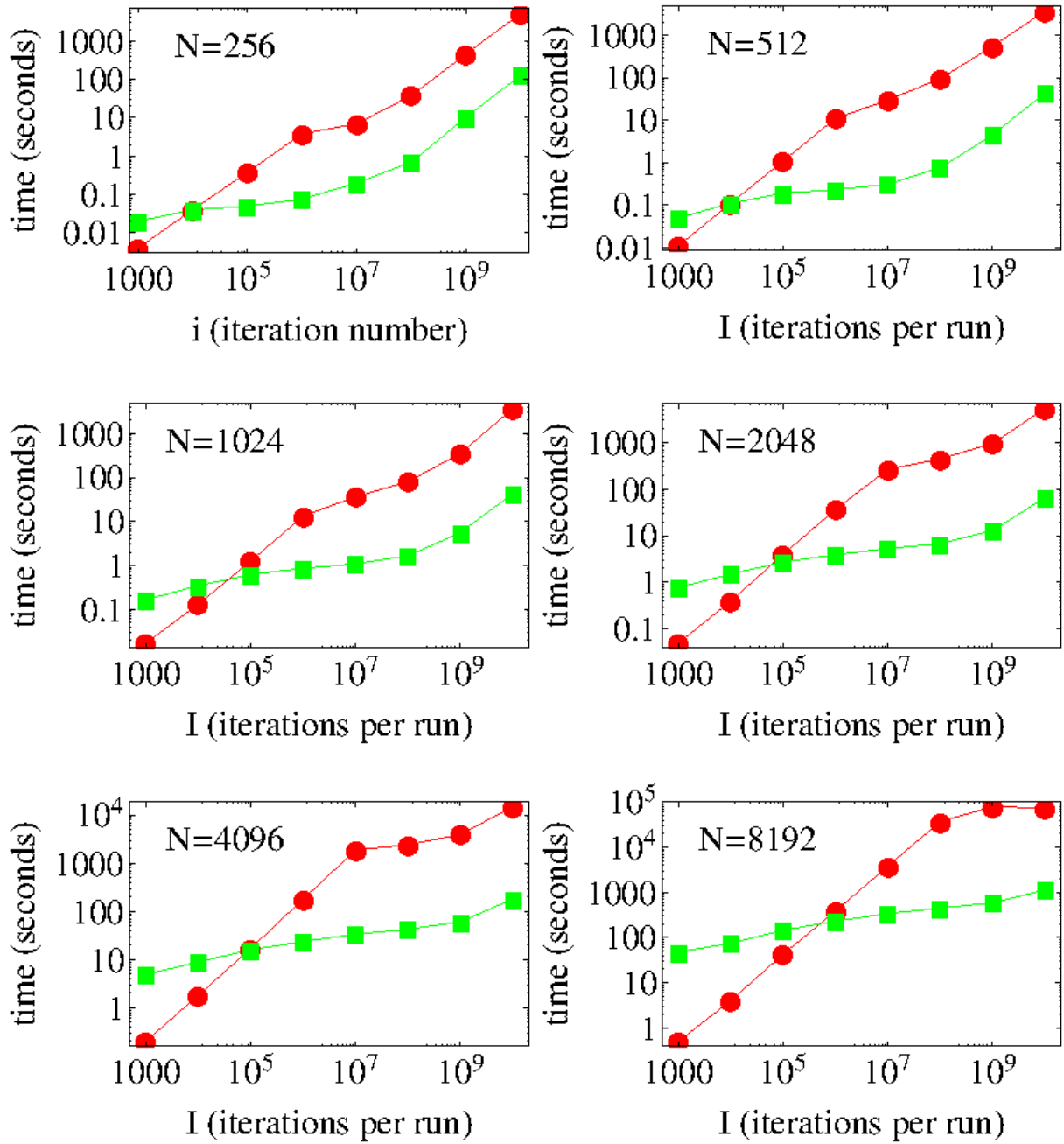

Figure 1: Run time versus number of iterations per run for random instances for non-GPU implementation (circles) and GPU implementation (squares).

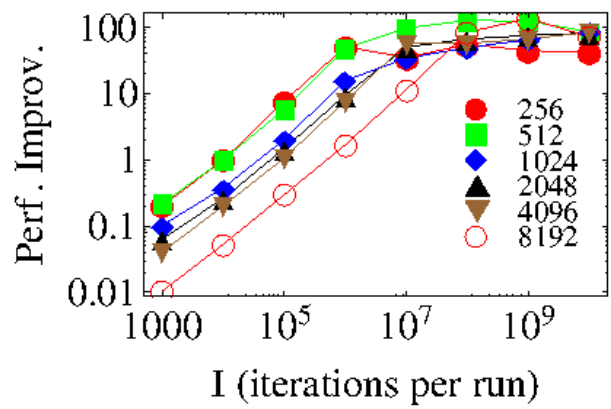

Figure 2: Performance improvement versus number of iterations per run